# Direct evidence for intermediate multiferroic phase in LiCuFe$_2$(VO$_4$)$_3$

Xiyu Chen,[1,#] Shuhan Zheng,[1,#] Meifeng Liu,[1,*] Tao Zou,[2] Wei Wang,[3] Keer Nie,[1] Fei Liu,[1] Yunlong Xie,[1] Min Zeng,[3] Xiuzhang Wang,[1] Hong Li,[1] Shuai Dong,[4,*] Jun-Ming Liu[1,5]

[1]*Institute for Advanced Materials, Hubei Normal University, Huangshi 435002, China*

[2]*Collaborative Innovation Center of Light Manipulations and Applications, Shangdong Normal University, Jinan 250358, China*

[3]*Institute for Advanced Materials, South China Academy of Advanced Optoelectronics, South China Normal University, Guangzhou 510006, China*

[4]*School of Physics, Southeast University, Nanjing 211189, China*

[5]*Laboratory of Solid State Microstructures, Nanjing University, Nanjing 210093, China*

**ABSTRACT:**

Magnetic susceptibility, specific heat, dielectric, and electric polarization of LiCuFe$_2$(VO$_4$)$_3$ have been investigated. Two sequential antiferromagnetic transitions at $T_{N1}$ ~ 9.95 K and $T_{N2}$ ~ 8.17 K are observed under zero magnetic field. While a dielectric peak at $T_{N1}$ is clearly identified, the measured pyroelectric current also exhibits a sharp peak at $T_{N1}$, implying the magnetically relevant ferroelectricity. Interestingly, another pyroelectric peak around $T_{N2}$ with opposite signal is observed, resulting in the disappearance of electric polarization below $T_{N2}$. Besides, the electric polarization is significantly suppressed in response to external magnetic field, evidencing remarkable magnetoelectric effect. These results suggest the essential relevance of the magnetic structure with the ferroelectricity in LiCuFe$_2$(VO$_4$)$_3$, deserving for further investigation of the underlying mechanism.

---

[#] The two authors contribute equally to this work.

[*] Corresponding authors. Email: lmfeng1107@hbnu.edu.cn (Meifeng Liu), sdong@seu.edu.cn (Shuai Dong)

## I. INTRODUCTION

The exploration and synthesis of novel multiferroics have been one of the major interests in condensed matter physics. In multiferroics, magnetism and ferroelectricity coexist, which allows the control of magnetization $M$ by electric field $E$ or electrical polarization $P$ by magnetic field $H$.[1-7] Therefore, multiferroics have huge potential applications in novel electronic devices, such as memories, sensors, etc.[4, 8] Depending on the origin of ferroelectricity, multiferroics are distinguished into two types.[1, 9] In type-I multiferroics, the magnetism and ferroelectricity have distinctly different origins, and the magnetoelectric (ME) coupling is usually weak. For type-II multiferroics, polarization $P$ is generated by specific magnetic structures, leading to strong ME effect. Now it is well known that the ferroelectricity in type-II multiferroics originates from spin-orbit coupling or spin-lattice coupling.[1, 3] For instance, the ferroelectricity in $TbMnO_3$ and $Ba_2XGe_2O_7$ ($X$ = Mn, Co, and Cu) is induced by inverse Dzyaloshinskii-Moriya interaction and spin-dependent $p$-$d$ hybridization respectively.[10-12] These materials have non-collinear spin orders and the spin-orbit coupling dominates. Besides, in materials such as $Ca_3CoMnO_6$, the $P$ is induced by exchange striction effect in collinear up-up-down-down spin order, which is dominated by spin-lattice coupling.[13, 14] It is noted that these couplings belong to high order quantum effect and the induced $P$ is usually small. Even worse, the magnetic ordering temperature in type-II multiferroics is usually low. Therefore currently available multiferroics are not suitable for application yet. Continuous efforts have been paid to explore novel materials, such as $CaMn_7O_{12}$, $BiMn_3Cr_4O_{12}$, etc.[15-18] However, the difficulties remain, and searching for new multiferroic is still a major issue in this area.

Frustrated quasi-one-dimensional (quasi-1D) antiferromagnets have received widespread interest due to the rich and largely unexplored physics, such as quantum criticality,[19, 20] $H$-induced magnetic ordering,[21] quantum spin liquid,[22] etc. Owing to the highly frustrated magnetic interactions, multiferroicity induced by non-collinear spin order is expected in quasi-1D magnetic systems. In fact, this issue has been explored in quasi-1D systems such as Cu-base quantum spin chain $LiCu_2O_2$[23] and $LiCuVO_4$,[24] pyroxene $NaFeGe_2O_6$[25] and $SrMnGe_2O_6$,[26] and zigzag chain $MnWO_4$.[27] In these materials, strong ME effect, such as $H$ switching *P*, has been extensively investigated. Considering the large number of quasi-1D materials, it is valuable to explore further the quasi-1D magnetic systems, addressing the multiferroicity.

From this perspective, a recently reported quasi-1D mixed spin chain triple vanadate LiCuFe$_2$(VO$_4$)$_3$ attracted research interests in the multiferroic community. LiCuFe$_2$(VO$_4$)$_3$ was first synthesized by Belik,[28] and its crystal structure is shown in Figure 1(a-b). Neighboring FeO$_6$ octahedra and CuO$_5$ triangular bipyramids are connected by edge-sharing and form quasi-1D Cu$^{2+}$-Fe$^{3+}$ mixed spin chains. According to earlier work, the frustration factor $f = |\theta_{CW}|/T_N$ for LiCuFe$_2$(VO$_4$)$_3$ is 12, where $\theta_{CW}$ is the Curie-Weiss constant and $T_N$ is the magnetic ordering temperature,[29] suggesting highly frustrated spin structure in LiCuFe$_2$(VO$_4$)$_3$. Besides, the first-principles calculation indicated the competition of ferromagnetic (FM) and antiferromagnetic (AFM) exchanges, which may lead to a non-collinear spin order and induce ferroelectricity.[30] Very recently, indirect evidences in terms of ferroelectricity and ME effect in LiCuFe$_2$(VO$_4$)$_3$ were reported,[30, 31] while direct measurement of polarization *P* is absent.

In this work, we present a comprehensive study of LiCuFe$_2$(VO$_4$)$_3$ including the magnetic susceptibility, specific heat, and electric polarization measurements. It will be revealed that LiCuFe$_2$(VO$_4$)$_3$ undergoes two successive AFM transitions at $T_{N1}$ ~ 9.95 K and $T_{N2}$ ~ 8.17 K, as evidenced by magnetic susceptibility and specific heat. Very interestingly, two sign-opposite pyroelectric peaks at $T_{N1}$ and $T_{N2}$, marking the generation and disappearing of polarization *P* respectively are identified. The ME effect in terms of varying polarization *P* under different *H* is also presented.

## II. EXPERIMENT

Polycrystalline samples of LiCuFe$_2$(VO$_4$)$_3$ were synthesized by the conventional solid-state reaction method. The stoichiometric mixtures of Li$_2$CO$_3$, CuO, Fe$_2$O$_3$, and V$_2$O$_5$ were sufficiently ground and fired at 600 °C for 24 hours in air. The obtained powders were reground and pelleted. Then, the resultant pellets were sintered at 640 °C for 72 hours in air again with three intermediate regrinding and pelleting processes. Eventually, the phase purity of powder samples were analyzed using an X-ray diffractometer (SmartLab Se, Rigaku) with Cu-$K\alpha$ radiation at room temperature ($T$).

The *dc* magnetic susceptibility $\chi(T)$ as a function of $T$ under different magnetic fields $H$ was measured using Physical Property Measurement System (PPMS DynaCool-9, Quantum Design). The specific heat ($C_p$) was also measured in PPMS DynaCool-9 based on the heat relaxation method.

For electrical measurement, Au electrodes were deposited on the top and bottom sides of the disk-like sample (3.25 mm in diameter and 0.3 mm in thickness). The measurements were carried out in PPMS which provides magnetic field and cryogenic environment. The dielectric constant $\varepsilon_r$ was measured by an LCR meter (Agilent E4980A). For pyroelectric current ($I_{pyro}$) measurement, the poling electric field $E_p$ was provided by Keithley 6517B electrometer. Before the measurement, the sample was cooled from 15 K to 2 K under $E_p = \pm\, 3.33$ kV/cm. When the sample temperature was cooled to 2 K, the poling electric field was removed and the sample was short-circuited for 30 minutes to remove the trapped charge. Then the pyroelectric current was recorded using the Keithley 6517B electrometer when raising the temperature. The $P$ is derived by integrating the pyroelectric current. It is noted that the magnetic field remained unchanged during the pyroelectric measurement.

## III. RESULTS

Figure 1c displays the measured X-ray spectrum of LiCuFe$_2$(VO$_4$)$_3$ powders. No stray peaks are identified, suggesting the high purity of LiCuFe$_2$(VO$_4$)$_3$. The diffraction patterns are further analyzed by Rietveld refinement.[32, 33] The space group of LiCuFe$_2$(VO$_4$)$_3$ is fitted to be triclinic $P$-1, and the refined structure parameters are $a = 8.1457(99)$ Å, $b = 9.8032(82)$ Å, $c = 6.6334(72)$ Å, $\alpha = 103.825(3)°$, $\beta = 102.361(0)°$, $\gamma = 106.992(4)°$. The results are consistent with earlier reports.[29-31] More detailed structural parameters are displayed in Table I.

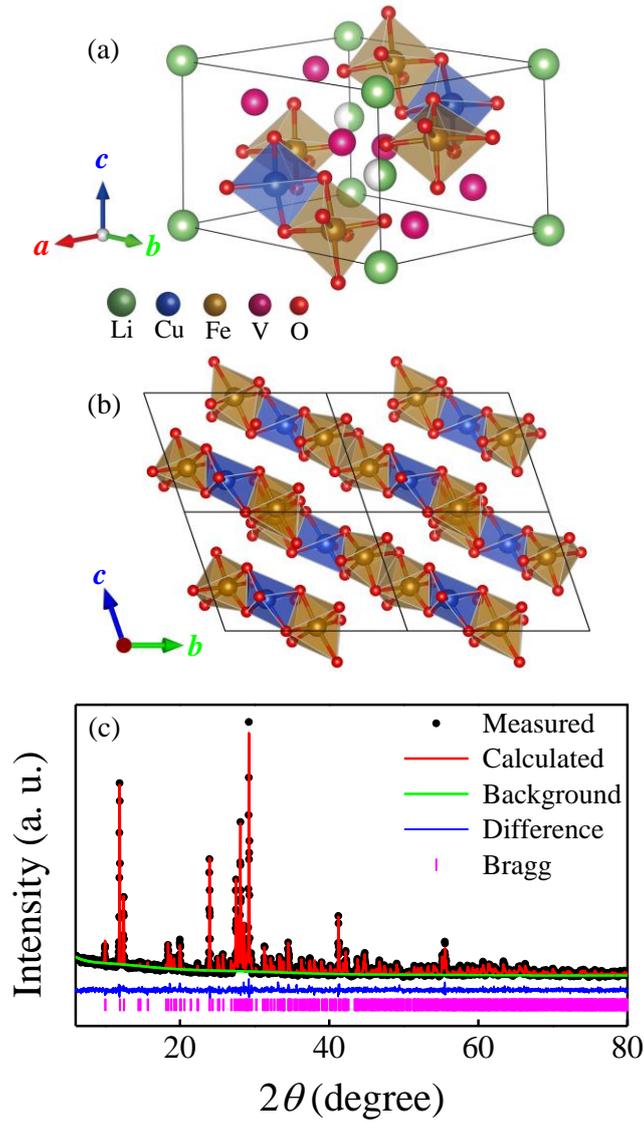

Figure 1. (a-b) Schematic of the crystal structure of LiCuFe$_2$(VO$_4$)$_3$. The unit cells are indicated by black boxes. (c) Rietveld refined powder X-ray spectrum of LiCuFe$_2$(VO$_4$)$_3$ at room temperature.

Table I. Refined structure information of LiCuFe$_2$(VO$_4$)$_3$ from powder X-ray diffraction at room $T$

| Atom (Wyck.) | x | y | z | Occ. | Site |
|---|---|---|---|---|---|
| Li1 | 0.0000 | 0.0000 | 0.0000 | 1.000 | 1a |
| Li2 | 0.0909(3) | 0.0743(0) | 0.5118(5) | 0.500 | 2i |
| Cu1 | 0.7780(0) | 0.2847(5) | 0.2667(9) | 1.000 | 2i |
| Fe1 | 0.4493(7) | 0.0969(3) | 0.3716(1) | 1.000 | 2i |
| Fe2 | 0.6990(2) | 0.5106(4) | 0.0343(0) | 1.000 | 2i |
| V1 | 0.6015(6) | 0.8320(9) | 0.1164(9) | 1.000 | 2i |
| V2 | 0.2257(7) | 0.3740(1) | 0.4082(9) | 1.000 | 2i |
| V3 | 0.1565(2) | 0.7601(0) | 0.2336(9) | 1.000 | 2i |
| O1 | 0.0205(0) | 0.2524(2) | 0.2885(2) | 1.000 | 2i |
| O2 | 0.5446(5) | -0.0985(6) | 0.3396(7) | 1.000 | 2i |
| O3 | 0.2700(4) | 0.4442(0) | 0.2607(5) | 1.000 | 2i |
| O4 | 0.3175(1) | 0.2325(2) | 0.4085(1) | 1.000 | 2i |
| O5 | 0.2700(7) | 0.7617(8) | 0.4707(5) | 1.000 | 2i |
| O6 | 0.5658(9) | 0.6527(0) | 0.0913(7) | 1.000 | 2i |
| O7 | 0.8239(9) | -0.0968(7) | 0.1619(6) | 1.000 | 2i |
| O8 | 0.5070(1) | 0.1539(7) | 0.1606(1) | 1.000 | 2i |
| O9 | 0.7816(0) | 0.3336(8) | -0.0018(1) | 1.000 | 2i |
| O10 | 0.7187(0) | 0.4874(9) | 0.3396(7) | 1.000 | 2i |
| O11 | 0.1921(9) | -0.0632(8) | 0.2165(8) | 1.000 | 2i |
| O12 | -0.0562(6) | 0.6712(0) | 0.1997(9) | 1.000 | 2i |

Space group: $P$-1, $a$ = 8.1457(99) Å, $b$ = 9.8032(82) Å, $c$ = 6.6334(72) Å, $\alpha$ = 103.825(3)°, $\beta$ = 102.361(0)°, $\gamma$ = 106.992(4)°, $R_p$ = 3.04%, $R_{wp}$ = 3.89%, $\chi^2$ = 1.121

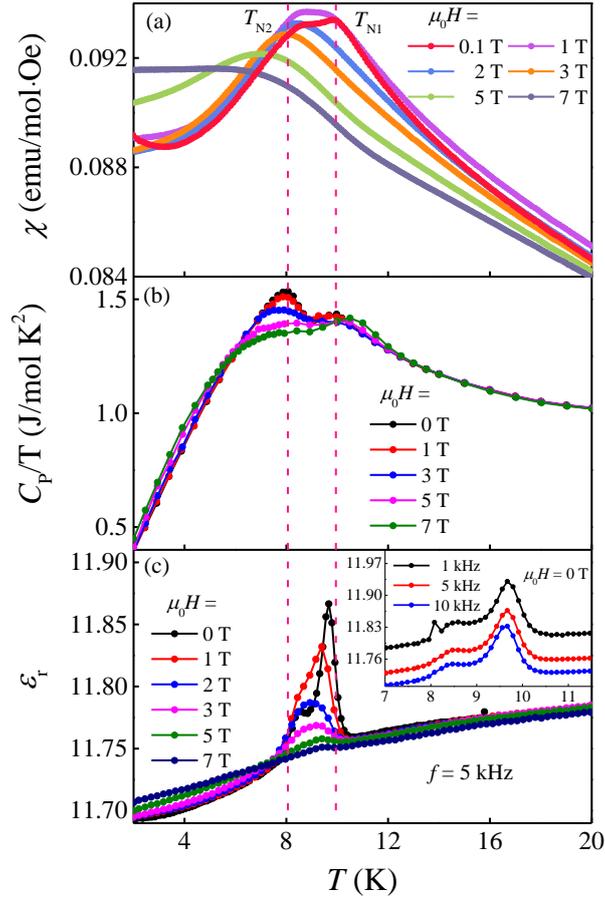

Figure 2. (a) $T$-dependent magnetic susceptibility under different $H$. (b) $T$-dependent $C_p/T$ of LiCuFe$_2$(VO$_4$)$_3$ under different $H$. (c) The dielectric constant $\varepsilon_r$ under different $H$ with $f$ = 5 kHz. The inset shows the $T$ dependence of $\varepsilon_r$ with various frequencies under zero $H$.

Figure 2a shows the magnetic susceptibility $\chi$ as a function of temperature measured at various magnetic field $H$. Two anomalies are found at $T_{N1}$ ~ 9.9 K and $T_{N2}$ ~ 8.6 K under $\mu_0 H$ = 0.1 T, indicating the onset of long-range magnetic ordering (LRO). With increasing $H$, the peaks of $\chi(T)$ slowly drift to lower $T$ and become broader. This suggests that the LRO of LiCuFe$_2$(VO$_4$)$_3$ is suppressed by $H$, which is a typical behavior of AFM system.[34, 35] To further check the magnetic transition of LiCuFe$_2$(VO$_4$)$_3$, we measured the specific heat $C_p$ of LiCuFe$_2$(VO$_4$)$_3$ under different $H$. Similarly, the $C_p/T$ of LiCuFe$_2$(VO$_4$)$_3$ under zero $H$ exhibits two serial peaks at $T_{N1}$ ~ 9.95 K and $T_{N2}$ ~ 8.17 K, as shown in Figure 2b. With increasing $H$, the peak at $T_{N2}$ drifts to lower $T$ and is suppressed. The specific heat measurement verifies that the AFM transitions of LiCuFe$_2$(VO$_4$)$_3$ are suppressed by $H$, which is consistent with $\chi(T)$ curves.

As to the electrical measurements, the dielectric constants $\varepsilon_r(T)$ is measured first, as presented in Figure 2c. The $\varepsilon_r(T)$ exhibits a sharp peak at $T_{N1}$ and a step-like anomaly at $T_{N2}$ under zero $H$, as shown in the inset of Figure 2c. These two anomalies do not shift with increasing measuring frequency, ruling out the possibility of artifactual signal from defects' relaxation. Then the magneto-dielectric effect is investigated. As shown in Figure 2c, the two anomalies are significantly suppressed and drift to lower $T$ with increasing $H$. And the anomalies are almost unrecognizable at $\mu_0 H$ = 7 T. These results are consistent with $\chi(T)$ and $C_p$ data. The remarkable magneto-dielectric effect hints the existence of ferroelectricity and remarkable ME effect. Actually, previous studies have claimed the multiferroicity of LiCuFe$_2$(VO$_4$)$_3$ by discussing the dielectric signals,[30, 31] but the direct ferroelectric measurements are still missing.

The pyroelectric current $I_{pyro}(T)$ is measured to reveal the electrical polarization of LiCuFe$_2$(VO$_4$)$_3$. Figure 3a shows $I_{pyro}(T)$ of LiCuFe$_2$(VO$_4$)$_3$ obtained at various heating rates (2, 4, and 6 K/min) under poling electric field $E_p$ = 3.33 kV/cm and zero magnetic field. Interestingly, two current peaks with opposite directions are observed at $T_{N1}$ and $T_{N2}$ respectively. The current peaks are un-shifted with various heating rates, indicating that the current signals are indeed from pyroelectric effect. Besides, the poling with a negative electric field indeed leads to a reversal of $I_{pyro}(T)$, as demonstrated in Figure 3a. Figure 3b presents the $P(T)$ curves obtained by integrating the $I_{pyro}(T)$. The $P(T)$ curves measured under $\pm E_p$ have similar magnitude with opposite directions, implying that the $P$ can be reversed and thus LiCuFe$_2$(VO$_4$)$_3$ is indeed a ferroelectric material. In short, through the combination of $\chi(T)$, $C_p$, $\varepsilon_r$, and $I_{pyro}(T)$ measurements, we exhibit the simultaneous existence of FE transition and AFM transitions, and confirm that LiCuFe$_2$(VO$_4$)$_3$ is a magnetically induced multiferroics (i.e., a type-II multiferroic one).

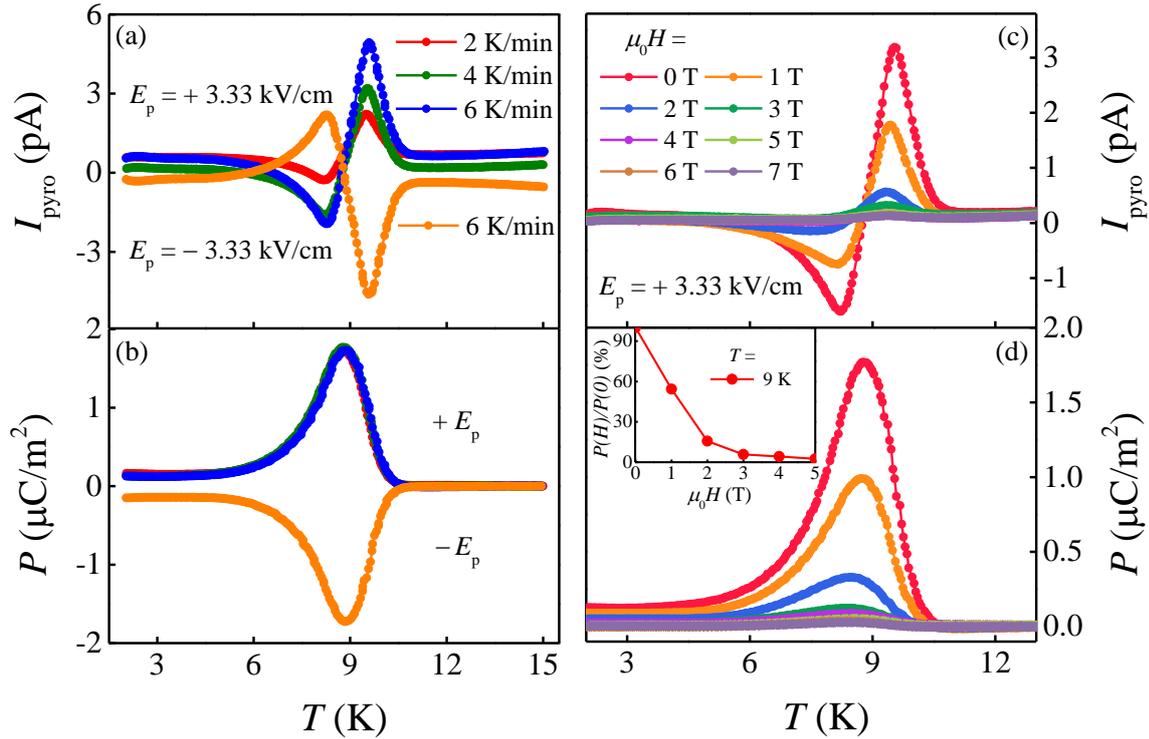

Figure 3. (a) The pyroelectric current of LiCuFe$_2$(VO$_4$)$_3$ measured at various heating rates: 2 K/min, 4 K/min, and 6 K/min under $E_p = \pm 3.33$ kV/cm and $\mu_0 H = 0$ T. (b) The $T$-dependent ferroelectric polarization derived by integrating the pyroelectric current relative to time under $E_p = \pm 3.33$ kV/cm and $\mu_0 H = 0$ T. (c) The pyroelectric current and (d) the corresponding ferroelectric polarization of LiCuFe$_2$(VO$_4$)$_3$ under various $H$ and $E_p = \pm 3.33$ kV/cm. Insert: ME effect at $T = 9$ K.

Then the ME effect of LiCuFe$_2$(VO$_4$)$_3$ can be demonstrated by measuring the $I_{pyro}(T)$ curves under different $H$, and the results are presented in Figure 3c. With increasing $H$, the two pyroelectric current peaks slightly shift to lower $T$ and the magnitudes decrease simultaneously. It is noted that the pyroelectric current is too weak to be recognized when $H$ is higher than 3 T. Figure 3d depicts the $P$ under different $H$, which is significantly suppressed by increasing $H$. Such behavior is widely observed in many type-II multiferroics, which further suggests that $LiCuFe_2(VO_4)_3$ is one of the cases.[13, 36]

## IV. DISCUSSION

According to above measurements, it is obvious that $LiCuFe_2(VO_4)_3$ belongs to the type-II multiferroics. On one hand, complicated magnetic interactions were argued to exist between Fe-Cu and Fe-Fe spins in $LiCuFe_2(VO_4)_3$,[29, 30] which might lead to frustrated spin states like non-collinear spin texture with incommensurate/commensurate period. On the other hand, the $P$ of $LiCuFe_2(VO_4)_3$ is highly susceptible to $H$, which can be attributed to the change of spin texture upon $H$. Both these characteristics are feature of type-II multiferroics. Besides these common characteristics, the multiferroicity in $LiCuFe_2(VO_4)_3$ is special for its two pyroelectric current peaks. There are two possibilities to explain this nontrivial behavior.

The first possibility is that these two peaks could be from two independent ferroelectric components with different transition points, like recently proposed "irreducible ferrielectricity".[37] In $LiCuFe_2(VO_4)_3$, there are two different magnetic ions ($Cu^{2+}$ and $Fe^{3+}$) and $Fe^{3+}$ ions have two nonequivalent positions. Therefore these $Cu^{2+}$ and $Fe^{3+}$ ions may order in different temperatures, and form two different sublattices which contribute to the net $P$ independently. Similar scenario can be referred to extensively researched $RMn_2O_5$[38-40] and $DyMnO_3$.[41, 42] In $RMn_2O_5$ two ferroelectric components are identified, which originate from $Mn^{3+}$-$Mn^{4+}$-$Mn^{3+}$ blocks and $R^{3+}$-$Mn^{4+}$-$R^{3+}$ blocks respectively.[43, 44] In $DyMnO_3$, the Mn sublattice and Dy sublattice orders in different temperatures, and contribute to the net polarization in opposite directions.[41, 42]

The second possibility is that the two current peaks correspond to sequential magnetic transitions. Before $T_{N1}$, the paramagnetic phase is nonpolar, and the AFM phase below $T_{N2}$ is also nonpolar. Only the AFM phase between $T_{N1}$ and $T_{N2}$ breaks the spatial inversion symmetry and generates the polarization. Such scenario of intermediate multiferroic phases have also been found in $MnWO_4$, $Ni_3V_2O_8$, and CuO.[27, 45-50] In these materials, multiple magnetic transitions take place at different temperature and the ferroelectricity only exists in the temperature windows of incommensurate magnetic structures.[27, 45-48]

According to above discussions, the complicated magnetic interaction in $LiCuFe_2(VO_4)_3$ makes the presence of two sequential magnetic transitions. Thus, the second scenario seems to be much more likely than the first one. To further verify this scenario, the sample was cooled down to 2 K without poling electric field, and then poled the sample from 2 K to 9 K with an electric field $E = \pm 3.33$ kV/cm. Then, the pyroelectric current of $LiCuFe_2(VO_4)_3$ was measured during the heating or cooling the samples. As shown in Figure 4a, the ferroelectric polarization $P$ only exists between $T_{N2}$ and $T_{N1}$, while it is negligible in other region. Besides, this $P$ is identical to the case with poling from 20 K to 9 K, as compared in Figure 4b. Hence, we are confidential that $LiCuFe_2(VO_4)_3$ is similar to CuO, which belongs to the family of intermediate multiferroic systems. Of course, the magnetic structure of $LiCuFe_2(VO_4)_3$ has not been resolved yet. In future, neutron diffraction experiments are desired to clarify this issue by solving the magnetic structures of $LiCuFe_2(VO_4)_3$ between $T_{N1}$ and $T_{N2}$, as well as below $T_{N2}$.

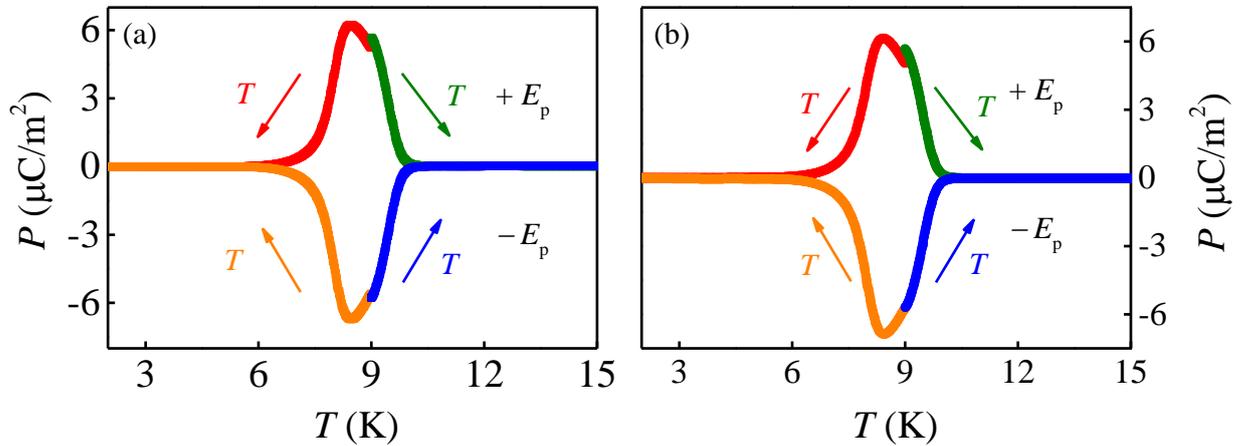

Figure 4. The ferroelectric polarization of $LiCuFe_2(VO_4)_3$. The corresponding pyroelectric current was measured when cooling the sample from 9 K to 2 K (red and orange) or heating the sample from 9 K to 15K (green and blue). (a) With a poling process from 2 K to 9 K. (b) With a poling process from 20 K to 9 K. The poling field is fixed as $E_p = \pm 3.33$ kV/cm.

Based on the second scenario, the magnetic and polar phase diagram for $LiCuFe_2(VO_4)_3$ is plotted in Figure 5, according to above measurements. It is paramagnetic (PM) and paraelectric (PE) above $T_{N1}$~9.95 K. When the $T$ decreases to $T_{N1}$, $LiCuFe_2(VO_4)_3$ undergoes the first AFM phase transition (so called AFM1 phase), and simultaneously a ferroelectric phase transition (FE phase) occurs at $T_{N1}$. As $T$ further decreases, another AFM transition (so called AFM2 phase) occurs at $T_{N2}$~8.17 K. The AFM2 phase transition leads to the disappearance of ferroelectricity. Therefore $LiCuFe_2(VO_4)_3$ is marked as paraelectric below $T_{N2}$. The applied magnetic field can slightly expand this multiferroic window. Under the magnetic field, the anomaly signal can slightly broad the ferroelectric phase region, which may be related to possible frustrated spin structure. Maybe the magnetic field can suppress the frustration, which enhances the ordering temperature slightly.

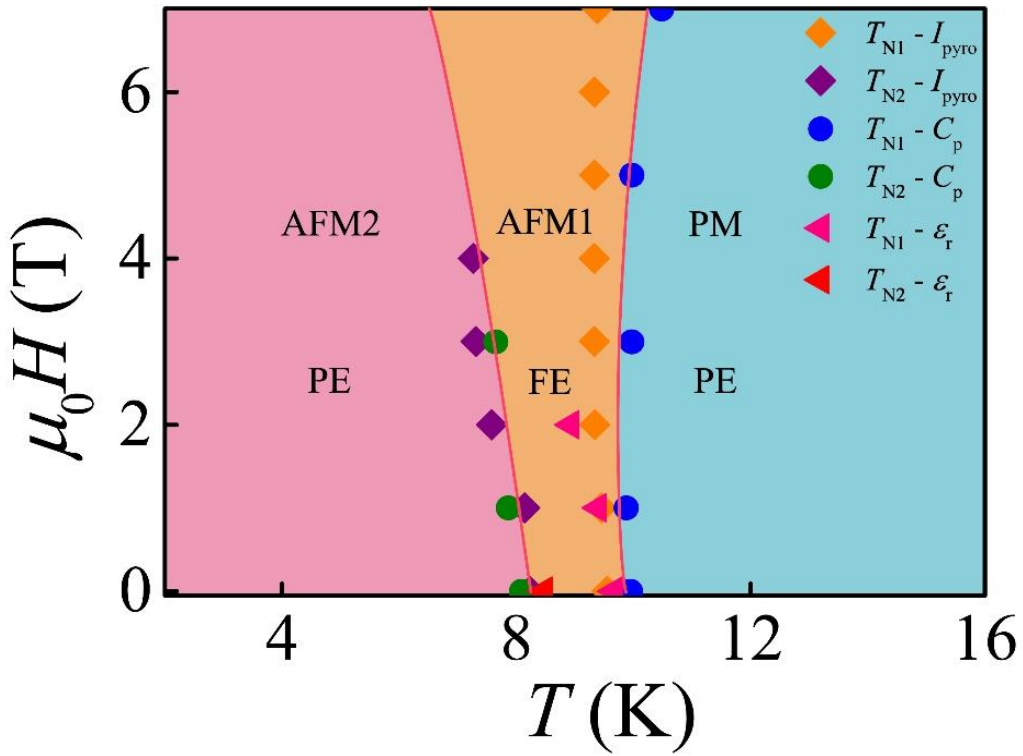

Figure 5. The magnetic and polar phase diagram for $LiCuFe_2(VO_4)_3$ determined by aforementioned measurements. The multiferroic phase exists only in the intermediate window.

## V. CONCLUSION

In summary, we have systematically investigated the magnetism and multiferroicity in mixed spin chain LiCuFe$_2$(VO$_4$)$_3$. Magnetic susceptibility and specific heat showed two adjacent AFM transitions at $T_{N1} \sim 9.95$ K and $T_{N2} \sim 8.17$ K. Pyroelectric current and dielectric measurements confirmed that the FE transition at $T_{N1}$. Interestingly, two sign-opposite pyroelectric peaks are observed at $T_{N1}$ and $T_{N2}$, giving rise to a reentrant paraelectric phase below $T_{N2}$. Moreover, the FE $P$ of LiCuFe$_2$(VO$_4$)$_3$ is very sensitive to magnetic fields, i.e., it exhibits significant ME effect. We suggest that the sequential magnetic transitions lead to this special multiferroic behavior, namely only the intermediate AFM1 phase can induce polarization. Further studies using single crystal samples and neutron experiments are encouraged to verify our proposal of multiferroicity in LiCuFe$_2$(VO$_4$)$_3$.

## ACKNOWLEDGMENTS

This work was supported by the National Natural Science Foundation of China (Grant Nos. 11834002, 12074111, 92163210, 11704109). The Research Project of Hubei Provincial Department of Education (Grant No. Q20202502).